# Effect of chromophore-chromophore electrostatic interactions in the NLO response of functionalized organic-inorganic sol-gel materials.+

J. Reyes-Esqueda[1,2], B. Darracq[1], J. García-Macedo[2], M. Canva[1], M. Blanchard-Desce[3], F. Chaput[4], K. Lahlil[4], J.P. Boilot[4], A. Brun[1] and Y. Lévy[1]

[1] Laboratoire Charles Fabry de l'Institut d'Optique
Centre National de la Recherche Scientifique – CNRS UMR 8501
Université d'Orsay-Paris XI, BP 147, 91403 Orsay Cedex, France

[2] Departamento de Estado Sólido, Instituto de Física – UNAM, Ciudad Universitaria, Del. Coyoacán. 04510, México, D.F., México.

[3] Département de Chimie
Centre National de la Recherche Scientifique – CNRS UMR 8640
Ecole Normale Supérieure, 24, rue Lhomond, 75231 Paris Cedex 05, France

[4] Laboratoire de Physique de la Matière Condensée
Centre National de la Recherche Scientifique – CNRS UMR 7643
École Polytechnique, 91128 Palaiseau, France

**Abstract.**

In the last years, important non-linear optical results on sol-gel and polymeric materials have been reported, with values comparable to those found in crystals. These new materials contain push-pull chromophores either incorporated as guest in a high $T_g$ polymeric matrix (doped polymers) or grafted onto the polymeric matrix. These systems present several advantages; however they require significant improvement at the molecular level - by designing optimized chromophores with very large molecular figure of merit, specific to each application targeted. Besides, it was recently stated in polymers that the chromophore-chromophore electrostatic interactions, which are dependent of chromophore concentration, have a strong effect into their non-linear optical properties. This has not been explored at all in sol-gel systems. In this work, the sol-gel route was used to prepare *hybrid organic-inorganic thin films* with different NLO chromophores grafted into the skeleton matrix. Combining a molecular engineering strategy for getting a larger molecular figure of merit and by controlling the intermolecular dipole-dipole interactions through both: the tuning of the push-pull chromophore concentration and the control of TEOS (Tetraethoxysilane) concentration, we have obtained a $r_{33}$ coefficient around 15 pm/V at 633 nm for the classical DR1 azo-chromophore and a $r_{33}$ around 50 pm/V at 831 nm for a new optimized chromophore structure.

+: Work supported by DGAPA UNAM IN 103199, and CONACYT 34582-E.

# 1. Introduction.

Recently, studies of non-linear optical effects has been focused into the field of organic polymeric materials because of its large potential efficiencies, its high malleability (allowing to this kind of materials to adopt a very wide variety of shapes), and its very low cost of fabrication. All these advantages have caused a great development of both active and passive optic components in the field of optical data storage and telecommunications[1,2,3,4,5,6].

The main feature of these new kind of non-linear materials is the incorporation of organic active molecules into a passive polymeric matrix. This incorporation is normally made in two ways. In one of them, the organic molecule is incorporated as a guest in the polymer (doped polymer) with the resulting loading limitation and aggregation problems. In the other one, the molecule is grafted or covalently attached into the structural skeleton of the matrix (functionalized polymer). This second way allows increasing the molecular concentration.

The organic active molecules normally incorporated into a polymeric matrix to obtain second order non-linear responses are those called push-pull chromophores, because due to their shape they have the presence in each extreme of a donor and an acceptor groups, respectively. This class of molecules have a distinctive figure of merit given as $\mu\beta/MW$, where $\mu$ is the dipole moment, $\beta$ is the first hyperpolarizability and $MW$ the chromophore molecular weight. The external electric orientation of these molecules allows obtain a non-centrosymmetric material that can be useful for applications in integrated optics, such as optical modulation[7]. According to this, most people have directed their efforts to improve the design of new chromophores with very high $\beta$-values in order to increase the materials nonlinear coefficient magnitude[8,9].

However, several authors have shown[6,8-10] that the chromophore concentration in a polymer matrix plays a role extremely important in the magnitude of these non-linear properties. Dalton and co-workers explained how chromophore-chromophore electrostatic interactions limit the optical non-linear maximum value that it can be achieved in electric-field poling experiments. Using London theory[11] for modeling intermolecular interactions between spherical shaped chromophores, the poling-induced electro-optic (E.O.) coefficient can be calculated in good approximation and it is proportional to:

$$r_{33} \propto \left(\mu\beta/MW\right)[1 - L^2(W/kT)]E_p, \qquad [1]$$

where $E_p$ is the poling field, $k$ is the Boltzmann constant, $T$ is the poling temperature, $L$ is the first order Langevin function ($L(x)=\coth(x)-1/x$) and $W$ is the potential energy of the chromophore dipolar moment that can be written as the sum of three terms, namely, the orientation force (interaction between permanent dipoles), the induction force (interaction between induced dipoles), and the dispersion force[11]:

$$W = \left(1/R^6\right)\left[2\mu^4/3kT + 2\mu^2\alpha + 3I\alpha^2/4\right], \qquad [2]$$

where $R$ is the average distance between chromophores, $\alpha$ is the linear polarizability, and $I$ is the ionization potential.

Although all this previous work supposes essentially spherical shaped particles, under some considerations it can be applied to Disperse Red 1 (DR1), i.e. non-spherical particles, as Dalton et al. show in their work[8].

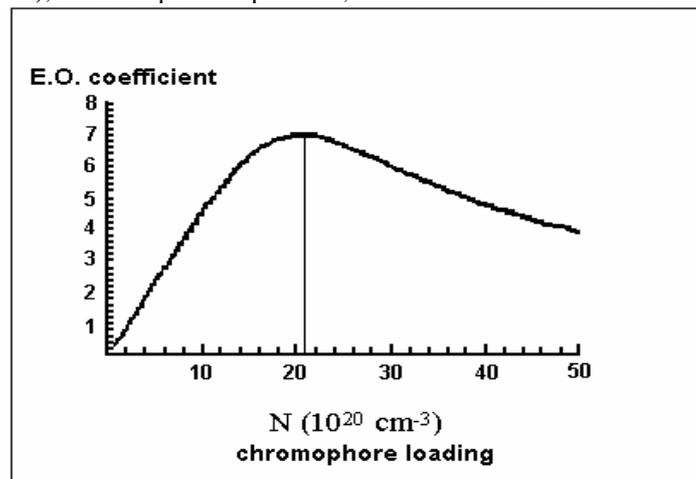

Figure 1. Theoretical variation of electro-optic coefficient, $r_{33}$, relative to chromophore loading (number density).

The most important conclusion of taking into account the intermolecular interactions is the theoretical variation of the electro-optic coefficient relative to chromophore loading as it is illustrated in Figure 1. This variation was experimentally corroborated in polymers by Dalton et al.[8], and as it is shown in the figure, the competition between chromophore-electric field and chromophore-chromophore electrostatic interactions leads to a maximum in the graphs of the electro-optic coefficient versus chromophore number density.

By the other way, parallel to the polymer development, another chemical way has shown non-linear results as important as those obtained in polymers. This way is the sol-gel process that was mainly used at the beginning in laser applications research [12,13] due to its very low fabrication temperature, which permit to introduce several organic compounds without decomposition[14,15].

As in polymers, non-linear optical (NLO) chromophores can be incorporated into the inorganic sol-gel matrix as a dopant or as side-chain units. This last way allowed to increase the molecular concentration and then, to enhance the temporal stability of the second-order optical non-linearities[16,17,18], but it also increased the aggregation problems. This fact means a limitation in the non-linear signal because of the intermolecular interactions introduced with the molecular aggregation.

For other purposes, like the development of photorefractive sol-gel materials, a charge-transporting unit had to be introduced as well. This unit was normally chosen to be the carbazole molecule (K) and it could also be attached to the sol-gel silicon skeleton. Collaterally, but as a very important consequence, the introduction of this Si-K unit meant the disappearance of aggregation problems, because the carbazole agents act as screeners for the electrostatic interaction between non-linear chomophores[16,19] [Figure 2]. This fact allowed increase the NLO molecule concentration furthermore, avoiding aggregation effects. This can happen for any other kind of molecules inside the sol-gel matrix, for example this occurs for the crosslinking agent TEOS (tetraethoxysilane), as it will be shown in this work for the classical NLO chromopore DR1.

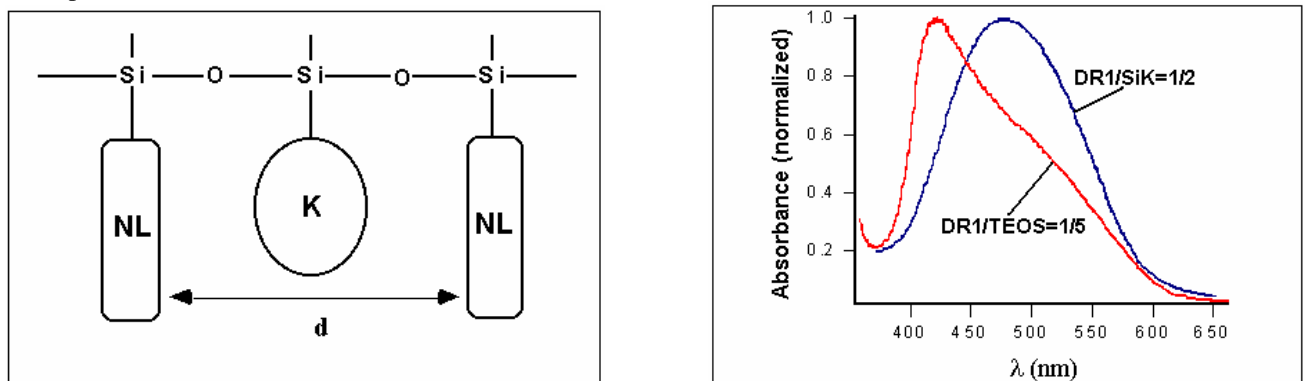

Figure 2. The carbazole units act as spacer between the NLO chromophore thus reducing their mutual interactions. Their effect on the absorption spectra is quite clear (see material preparation section for nomenclature).

In this work, the sol-gel method was used to prepare *hybrid organic-inorganic thin films* with different NLO chromophores grafted into the skeleton matrix. We will show how by combining a molecular engineering strategy in order to get larger molecular figures of merit, and controlling the intermolecular dipole-dipole interactions, we obtained a $r_{33}$ coefficient around 15 pm/V at 633 nm with the classical DR1 azo-chromophore. The dipole-dipole interactions are controlled via both tuning the push-pull chromophore concentration (by incorporation of screening carbazole moieties in high concentration), and controlling the TEOS (Tetraethoxysilane) concentration. Similarly, we obtain a $r_{33}$ around 50 pm/V at 831 nm (out of the main absorption) for a new optimized chromophore structure. For another new molecule with a large figure of merit, $\mu\beta$, and a high polarizability, $\alpha$, we also show how the molecular interactions can inhibit completely the second order response of the correspondent functionalized sol-gel material.

## 2. Experimental.

The molecular engineering and the sample preparation have been previously reported[16,20,21]. Two new push-pull chromophores, thereafter referred as MM42 and MM52, were prepared from its corresponding polyenal via a direct Konevenagel condensation, followed by acidic deprotection of the THP protecting group[20]. The polyenals used in the synthesis were obtained from the generic protected aldehyde via a protocol based upon vinylic extensions. The crude compounds were purified by column chromatography on silica gel and characterized by NMR and mass spectroscopy. The molecular precursors were prepared by reacting carbazole-9-carbonyle chloride with 3-aminopropyltriethoxysilane, yielding the charge-transporter precursor SiK, and by reacting the chromophores bearing free hydroxil groups with 3-isocyanatopropyltriethoxysilane yielding the push-pull NLO precursors SiDR1, SiMM42 and SiMM52[16]. To prepare coating solutions, the modified silanes in this way were copolymerized with tetraethoxysilane (TEOS) used as crosslinking agent as follows: The hydrolysis of these groups was made by adding acidulated water (pH=1) to the solution using tetrahidrofuran (THF) as a common solvent. The molar ratio for this is (SiDR1+SiK+TEOS):H$_2$O:THF=1:28:47[16]. After 2 hrs of hydrolysis, a little amount of pyridine is incorporated in order to neutralize the acidity of the solution and then to promote the condensation reactions[16]. The solution is filtered (45 µm) before depositing the samples by spin-coating. The TEOS concentration was controlled with respect to the SiDR1+SiK concentration. In this case, this last was fixed in the ratio SiDR1:SiK=1:5 and the TEOS concentration was varied into the range (SiDR1+SiK):TEOS=1:1, 3:1, 4:1, 6:1, 10:1 and ∞:1 (this last means no pure TEOS at all). In the same way, the NLO chromophore concentration was modulated with respect to the SiK concentration. The relative composition to the TEOS agent was fixed as (SiNLO+SiK):TEOS=6:1, and the SiNLO:SiK compositions were 1:2, 1:3, 1:4, 1:5, 1:7, 1:10 and 1:20 for DR1 and MM52 molecules, and 1:2, 1:10 and 1:20 for the MM42 one. Films of around 1 μm in thickness were made by spin-coating on several substrates. To study linear absorption, the films were deposited on glass substrates. To study Second Harmonic Generation (SHG) and to measure the electro-optic coefficient $r_{33}$, they were deposited on indium tin oxide (ITO)-covered glass substrates. After the preparation, the films were dried at 120 ºC during 1 hr under vacuum, and then the solvent was evaporated by a thermal treatment at 120 ºC during 15 hr. To measure the highest electro-optic coefficient an additional metallic electrode was deposited on the film sample side, immediately after the poling process.

The linear absorption was measured using an OMA spectrograph. The nonlinear optical properties were induced by the orientation of the NLO groups in the sol-gel film by a single-point Corona poling technique (needle-surface distance=12 mm, voltage=+6 kV, poling temperature=120 ºC). The SHG signal was carried out with a diode-pumped Nd:YAG laser operating at 1064 nm. The relaxation behavior was studied by monitoring the decay of the SHG signal as a function of time, at 120 ºC without applied field. The Pockels coefficient was measured by the simple reflection technique[17] at 633 nm and 831 nm using a He-Ne laser and a diode laser, respectively. The absorption spectra for each NLO chromophore after thin film preparation are shown in figure 3. The microscopic features for these molecules are shown in table 1.

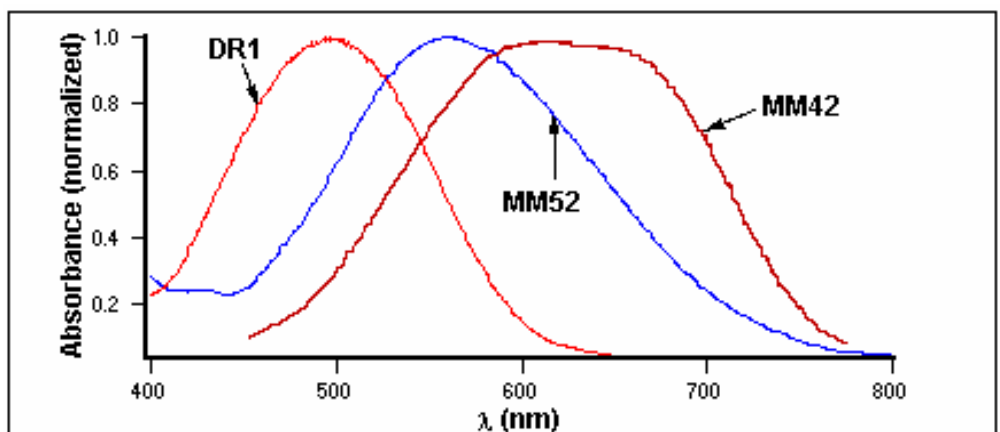

Figure 3. NLO chromophores (DR1, MM42 and MM52) doped sol-gel absorption spectra.

|      | $\alpha(0)^{[EOAM]}(10^{-24}$ esu$)$ | $\beta(0)^{[EFISH]}(10^{-30}$ esu$)$ | $\mu^*$(D) |
|------|---|---|---|
| **DR1**  | 38 | 47  | 8.7 |
| **MM52** | 50 | 320 | 10  |
| **MM42** | 75 | 300 | 8.6 |

Table 1. Microscopic features of NLO molecules studied in a poor polar environment.
*- Calculated using Guggenheim[22]-Debye[23] formula.

### 3. Results and discussion.

**3.1. Linear Absorption.**

The absorption spectra for the different TEOS concentrations showed that the amount of TEOS inside the films did not affect the DR1 concentration as long as the absorbance was the same for all the samples studied.

In the case of DR1 concentration, as this was increased, a blue shift of the maximum absorption was observed. This effect is shown in figure 4 for the DR1 normalized spectra. This blue shift can be attributed to dipole-dipole interactions leading to multimeric association of chromophores[24]. We can explain this shift as follows: the light-induced dipolar field of every chromophore, at any given chromophore site (each one treated as a point dipole), points always parallel, but opposite to, the external field, correspondent to the illuminating circularly-polarized-light, that is parallel to the surface of the film. This means that the local field, at any chromophore site, is always smaller than the external field[25]. As a consequence, the frecuencies of the normal modes of the system are always blue-shifted with respect to the resonance frequency of an isolated chromophore[25]. The shift increases when the chromophore-concentration increases since the packing of the light-induced dipoles becomes more compact as the concentration becomes higher, then becoming smaller the local field. This minimum-energy packing is a kind of H-aggregation with parallel chains of induced-dipoles stacked one on top of the other (figure 5).

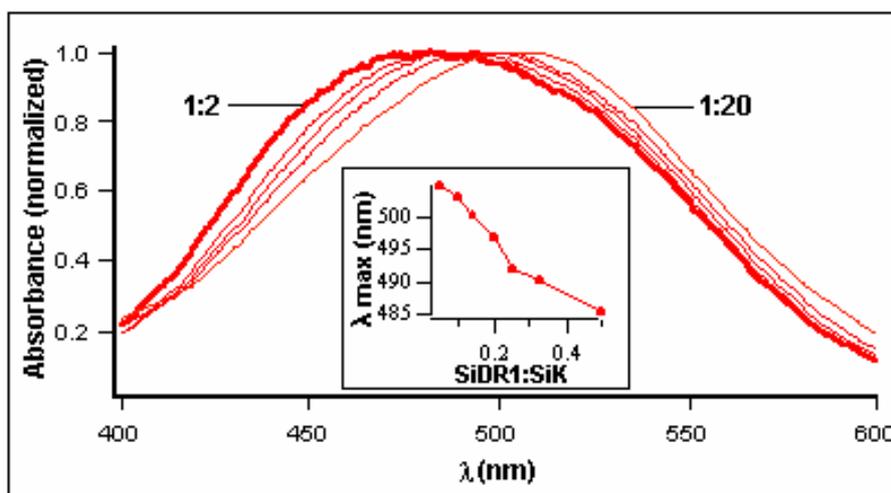

Figure 4. Blue shift of the maximum in the DR1 absorption spectra as the concentration was increased.

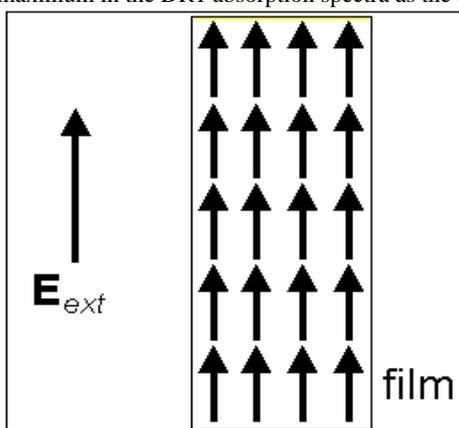

Figure 5. Transversal view of the packing of the light-induced dipoles into the sample for a external field parallel to the surface of the film.

In the case of MM42 molecule, the absorption shows a strong aggregation effect even for low concentrations, besides of the same blue shift as concentration increases. This behavior can be observed in figure 6 and it can be related to dipolar interactions and to the high polarizability observed in this molecule, as it will be explained below. Similar to DR1, the blue shift indicates how much the local field becomes smaller than the external one, when increasing concentration.

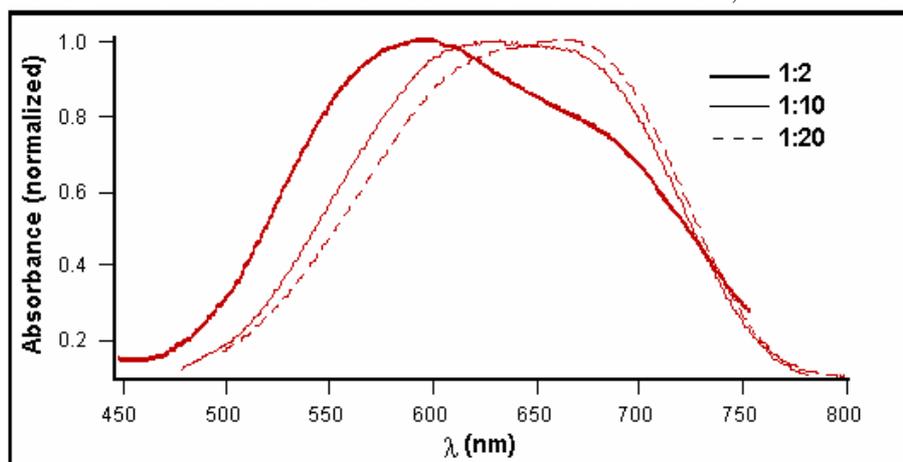

Figure 6. Absorption spectra for MM42 molecule.

**3.2. Second Harmonic Generation.**

### 3.2.1. Poling Process.

The stabilization time for the SH signal could not be related to the TEOS amount into the samples, besides the percentage of the SH signal decrement during the cooling process into the poling was constant (around 13%), meaning that NLO molecular relaxation was not affected by TEOS concentration. However, the SH maximum and final signals reached during poling indicates that too much TEOS (1:1 or 3:1) or no TEOS at all ($\infty$:1) affects the poling efficiency. This can be interpreted in terms of the intermolecular interactions because TEOS molecule is electrically neutral and it can screen the NLO molecular interactions, then when there is no TEOS at all, there is no way to screen them. In the same way, when there is too much TEOS, and the DR1 concentration remains the same, there is a very high molecular density into the film, and this fact inhibits the chromophore mobility induced during the poling. All this means that there is a suitable TEOS concentration for an optimal orientation.

By the other side, in Figure 7 a comparison among the poling dynamics of the different DR1 samples studied is shown. The 1:2 sample showed the highest SH signal at the beginning of the poling process, but after some minutes a clear decrement was observed. It can be attributed to the high chromophore concentration. Once the chromophore mobility is increased with temperature, and the external field is applied, there is an increment of chromophore-chromophore interactions to stabilize the system[26-29], and this is the cause of the signal lost. Last statement was verified in the 1:3 and 1:4 sample poling dynamics (the last one is not shown in the figure), since the same effect was observed but at minor scale. Conversely, this effect is not observed into the 1:5, 1:10 and 1:20 dynamics at all. But it has to be stated that the time to achieve the same signal was longer, and in the case of the 1:20 sample, it was never reached. The poling time was around two and a half-hours for all the samples. Moreover, once the temperature was down, the signal losses for the 1:10 and 1:20 samples were dramatic.

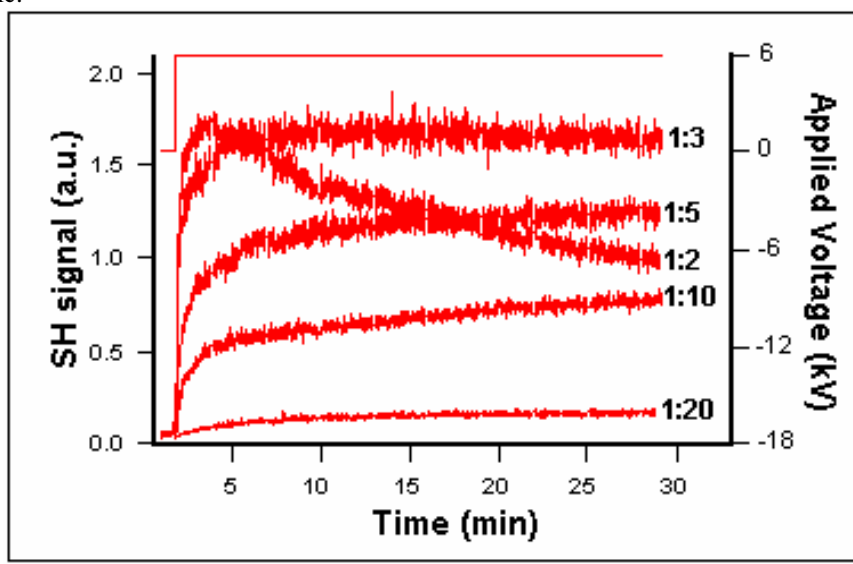

Figure 7. Comparison among the poling samples dynamics with different DR1 concentrations.

For the MM52 molecule the results found were similar, although the optimal concentration was different. For example, the effect observed for the DR1 1:2 concentration was already observed for the MM52 1:3 concentration, i.e., at lower chromophore concentration. This can be explained in terms of the microscopic characteristics for each molecule by looking at Table 1 where those values are shown. We can see that MM52 molecule has a $\mu\beta$ of around 8 times larger than that of DR1, then we can expect more pronounced molecular interactions at lower concentrations for the MM52 molecule.

For the MM42 molecule the results were extreme: the SH signal for the 1:2 concentration was zero, and for the 1:10 and 1:20 it was very small. This can be explained by dipolar interactions, that are very intense in this case. By looking at table 1, we can see that MM42 microscopic parameters are almost the same than those for MM52, however, the polarizability is 1.5 bigger, and it is two times that of DR1. By taking the theoretical model for the non-linearity given above, it can be observed in figure 8 how the polarizability increment induces the decrement of the non-linearity (reached through the poling process) because of stronger molecular interactions. When the polarizability is four times that of DR1, the dipersion force dominates the orientation and the induction forces.

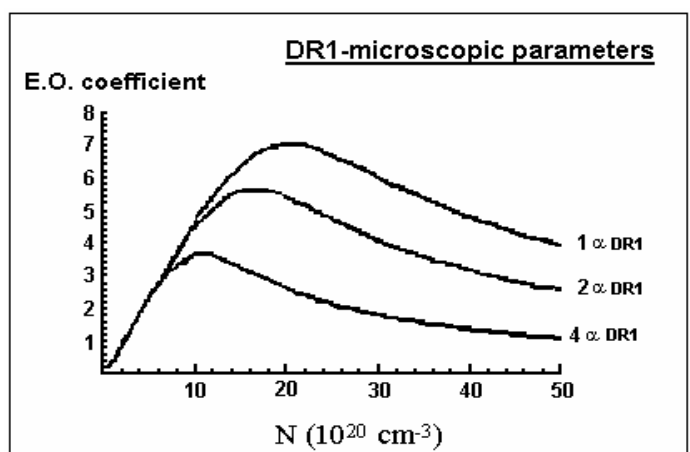

Figure 8. Polarizability influence on the non-linearity (E.O. coefficient) of the system.

### 3.2.2. Relaxation Process.

To test the effect of chromophore-chromophore interactions at higher concentrations we had to study the relaxation process. This meant to turn off the external field at 120 ºC. The results obtained with variation of TEOS concentration were surprising. As it can be remarked in figure 9, the fastest thermal relaxation was observed for the (SiDR1+SiK):TEOS=1:1 concentration, i.e. for the sample were the crosslinking agent concentration was the highest. Similarly, in the same figure it can be noted that the highest percentage of SH signal relaxation after 1 week was presented by the highest TEOS concentration: 1:1. In the reverse way, this percentage was zero for the samples where there was not pure TEOS at all (∞:1). This behavior is opposite to the expected for the crooslinking agent: we supposed at the beginning that by increasing the TEOS amount into the material, thermal and temporal relaxations would be smaller inside it. Besides, it has to be taken into account the previous poling result stating that it is necessary an optimum TEOS concentration in order to screen the NLO molecular interactions. All this together allows to state that for this kind of materials, a TEOS concentration around (SiDR1+SiK):TEOS=6:1 is the most suitable for an optimal poling, and for obtaining the highest electro-optic coefficient.

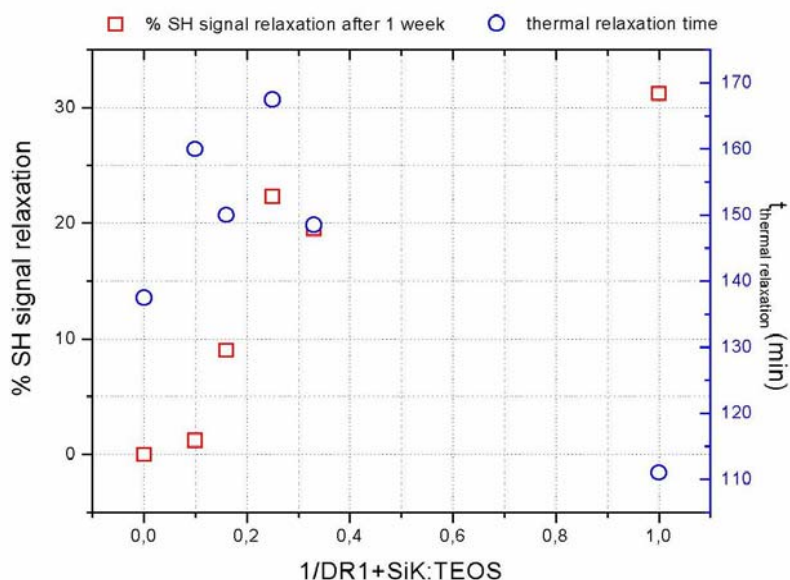

Figure 9. Thermal relaxation time and SH percent relaxation as function of TEOS concentration.

In the case of DR1 molecule, as it was expected, the higher the concentration the faster the signal decreases because of the dipole pairing that destroys the non-centersymmetry previously induced with the poling. This is shown in figure 10.

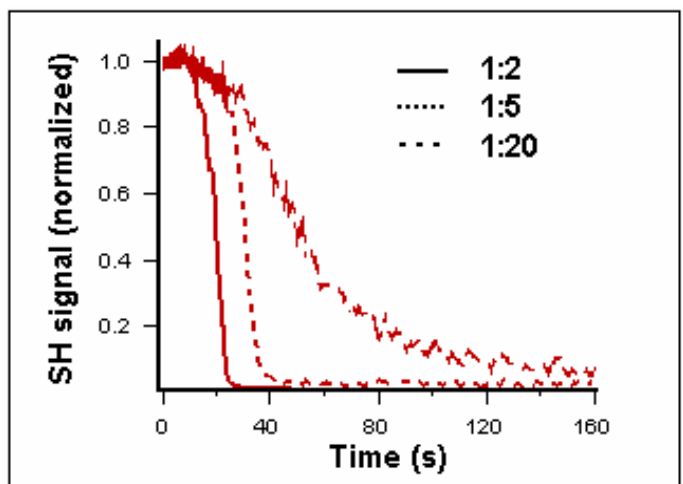

Figure 10. SHG signal relaxation when the poling field is turned off.

### 3.3. Electro-optic Coefficient.

To finish with the dipolar interactions effect into the non-linear properties of this kind of materials, the curves for Pockels coefficient against concentration obtained for the DR1 molecule at 633 and 831 nm are shown in figure 11. It is clear that the shape of them is the same as the theoretical one and that obtained experimentally for Harper and collaborators in polymers[8]. Similar to their work, in the sol-gel matrix an optimal NLO molecular concentration can be found depending of what wavelength is used, although in this case the poling dynamic has to be taken into account to choose it. The highest coefficients found with this poling dynamic (6 kV at 120 ºC around two and a half-hours) were found to be 13.4 pm/V at 633 nm for the DR1 1:2 concentration and 6 pm/V at 831 nm for the DR1 1:3 concentration.

In the case of the MM52 molecule, after analyzing its poling dynamics for all the concentrations studied, we decided to measure the Pockels coefficient for the 1:5 concentration finding an $r_{33}$ coefficient of 48 pm/V at 831 nm. This coefficient is the largest reported in sol-gel materials. But, for the MM42, due to its poor induced orientation, the measured coefficients were very small: 0, 2.5 and 0.5 pm/V at 831 nm for the MM42 1:2, 1:10 and 1:20 concentrations, respectively.

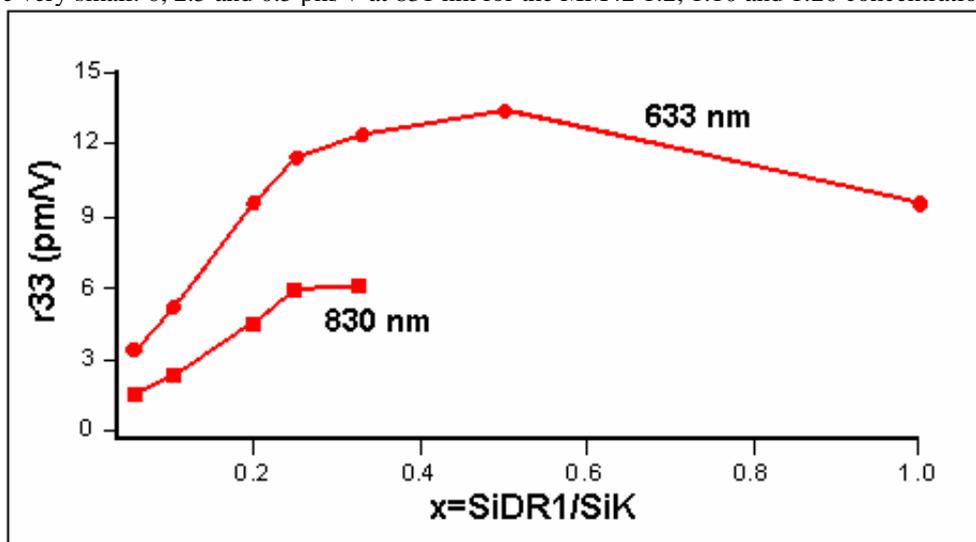

Figure 11. Pockels coefficient against concentration for the DR1 molecule at 633 (λ) and 831 (ν) nm.

### 4. Conclusions.

As it was shown in this work, when using a sol-gel matrix with a NLO functionalized chromophore, SiK units with carbazole groups at high concentration are very useful to avoid dipolar chromophore interactions and consequent aggregation problems. A suitable crosslinking agent (TEOS) concentration can be found in order to screen the molecular interactions and also to avoid, as much as possible, the orientational relaxation. Surprisingly, this TEOS concentration has to be low. As a general result, it is possible to control chromophore loading with respect to SiK units by fixing the TEOS concentration. The results obtained are similar to those found in polymeric materials, particularly the shape of the electro-optic coefficient curve vs. concentration is the same as that found for Harper *et al*. An important result is that an optimal chromophore concentration can be found for each used wavelength, after taking into account the poling process and the specifically targeted figure of merit. In this sense, another important result is that the figure of merit has to include the polarizability since as higher it is, the smaller is the non-linearity obtained, as it has been shown in this work for a new molecule with a polarizability two times that of classical DR1. To get this we propose to improve not only the hyperpolarizability $\beta$ of the molecule, but also the parameter $\mu/\alpha$, which is given for the ratio of the terms related with the orientation and the induction forces. Finally, a very high $r_{33}$ coefficient of 48 pm/V at 831 nm has been found for a molecule with a $\mu\beta$ eight times larger than that of DR1, suitable for electro-optics based applications.

## Acknowledgements

We acknowledge illuminating discussions with Cecilia Noguez.

[24] M. Kasha. "*Energy transfer mechanisms and the molecular exciton model for molecular aggregates*". Radiation Research. **20**, 55 (1963).

[25] C. Noguez, R. Barrera. "*Disorder effects on the effective dielectric response of a linear chain of polarizable spheres*". Physica A **211**, 399-410 (1994).